\documentstyle[sprocl]{article}
\input{psfig}   
\def\be{\begin{equation}}
\def\ee{\end{equation}}
\def\bea{\begin{eqnarray}}
\def\eea{\end{eqnarray}}

\def\be{\begin{equation}}
\def\ee{\end{equation}}
\def\ba{\begin{eqnarray}}
\def\ea{\end{eqnarray}}

\begin{document}
\title{Astrophysical Effects of Extreme Gravitational Lensing Events}
\author{Yun Wang and Edwin L. Turner}
\address{Princeton University Observatory, Peyton Hall, Princeton,\\
NJ 08544, USA}
\maketitle\abstracts{		
Every astrophysical object (dark or not) is a gravitational
lens, as well as a receiver/observer of the light from 
sources lensed by other objects in its neighborhood. For a 
given pair of source and lens, there is a thin on-axis 
tubelike volume behind the lens in which the radiation flux
from the source is greatly increased due to gravitational 
lensing. Any objects which pass through such a thin tube or beam 
will experience strong bursts of radiation, i.e., Extreme 
Gravitational Lensing Events (EGLEs). We have studied the 
physics and statistics of EGLEs.
EGLEs may have interesting astrophysical effects, such as 
the destruction of dust grains, ignition of masers, etc. 
Here we illustrate the possible astrophysical effects of 
EGLEs with one specific example, the destruction of dust 
grains in globular clusters.}

We propose a new way of looking at gravitational lensing by
noting that, 
for any given pair of source and lens, there is a thin on-axis 
tubelike volumn behind the lens in which the radiation flux
from the source is greatly increased due to gravitational 
lensing. Any objects which pass through such a thin tube or beam
will experience strong bursts of radiation, i.e., Extreme 
Gravitational Lensing Events (EGLEs).

Inside an EGLE beam, the flux from the source is {\it greater} than
a given value $f$; the larger $f$, the thinner the EGLE beam.
The characteristic cross section of an EGLE beam is given by
\be
\label{eq:xectpmax}
\pi r_{EGLE}^2(f) = \frac{8\pi R_{\rm S}}{27 D_{\rm ds}^3}
 \left( \frac{L_{\rm S}}{4\pi f}\right)^2,
\ee
where $R_{\rm S}=2GM/c^2$ is the Schwarzschild radius of the lens
of mass $M$, $D_{\rm ds}$ is the distance between the lens and the
source, and $L_{\rm S}$ is the luminosity of the source.

For a point source, the EGLE beam is infinitely long; it tapers
off at infinity. For a finite source, the EGLE beam ends at
a distance behind the lens, where the maximum magnification
of the source is just enough to bring the unlensed flux up to
$f$, the minimum flux inside the EGLE beam. Let us define
a dimensionless parameter
\be
\alpha(f) \equiv \frac{8R_{\rm S} D_{\rm c}}{\rho^2} \left(\frac{L_{\rm S}}
{4\pi D_{\rm c}^2 f}\right)^2,
\ee
where $D_{\rm c}$ is the maximum separation between the lens and the
source, and $\rho$ is the physical radius of the source.
$\alpha(f)$ measures the maximum magnification of the source relative 
to the flux $f$. $\alpha \rightarrow \infty$ is the point source limit;
the length of the EGLE beam increases with $\alpha$.

EGLEs may have interesting astrophysical effects.
For an astrophysical system with a population of sources and a
population of lenses, space is crisscrossed by a complex network 
of very bright but narrow beams of light and other forms of radiation.
Fig.1 shows a cartoon of the network of EGLE beams for five
sources (larger dots) and 18 lenses (smaller dots); the EGLE beams
produced by the center source are indicated by dotted lines to
illustrate the orientation of the EGLE beams.
\begin{figure} 
\begin{center}
\psfig{figure=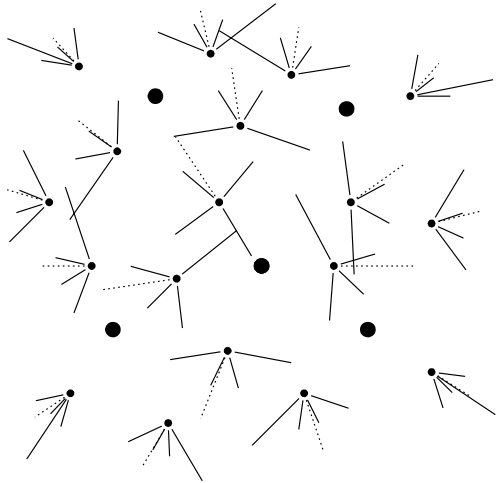,height=3.5in} 
\caption{A cartoon of the network of EGLE beams.}
\end{center}
%\label{fig:radish}}
\end{figure}
Without considering EGLE, the hot regions in this system are just
the several small spheres around each source.
Taking into consideration EGLE, each source
produces a thin hot beam which comes out of each lens and points away 
from the source. Additional hot regions which are 
thin beams coming out of each lens are thus produced.

In a globular cluster, there are several very bright and hot x-ray sources 
(believed to be accreting neutron stars)
and a large number of stars which act as lensing objects. 
In a typical globular cluster there would be millions of EGLE beams, 
one produced by each ordinary star acting on x-rays from each of the
cluster's sources.
These EGLE regions form a complex network of very hot and narrow
beams. Dust grains drift about in this system.
For dust grains to be destroyed by EGLE, the following two conditions 
are sufficient: (1) One EGLE heats up a dust grain to sufficiently high 
temperature for a sufficient amount of time to destroy the dust grain;
(2) Time between two EGLEs is less than the typical lifetime of a dust grain 
in the absence of EGLE. 
Both these conditions are satisfied in a typical globular cluster.
Whenever a dust grain crosses into an EGLE beam, it will be evaporated
and hence destroyed.
EGLE may in fact explain why no dust has been observed in globular clusters;
they could all have been destroyed by EGLE.

In conclusion, we note that
extreme gravitational lensing can be a source of perturbation in 
astrophysical systems of all scales; it may eventually provide the 
simplest explanation for some 
unexplained astrophysical phenomena.
Extreme Gravitational Lensing Events can be very dramatic when the
source is extremely bright (for example, a gamma-ray burst), and the
lens is very massive (for example, a giant black hole).
EGLE could provide the trigger for threshold phenomena;
it means that objects in interstellar space are subjected to
much larger variations in their radiation environment than
had been realized previously.

\section*{Acknowledgments} 
We gratefully acknowledge support from NSF grant AST94-19400.

\end{document}